\makeatletter
\declare@file@substitution{revtex4-1.cls}{revtex4-2.cls}
\makeatother
\documentclass[twocolumn]{aastex631}
\usepackage{listings}
\usepackage{subfigure}
\usepackage{xcolor}
\usepackage{xcolor, soul}
\sethlcolor{yellow}
\usepackage{enumitem}
\hypersetup{backref=true, pagebackref=true, hyperindex=true, colorlinks=true,                
	breaklinks=true, urlcolor= magenta, linkcolor= black, bookmarks=true,                 
	bookmarksopen=false, filecolor=cyan, citecolor=blue, linkbordercolor=blue}
\usepackage{hyperref}

\shorttitle{multi-planetary systems and Gravitational Microlensing}
\shortauthors{Fatheddin and Sajadian}

\begin{document}
\title{Detecting Multi-Planetary Systems with Gravitational Microlensing and the \textit{Roman} Space Telescope}

\author[0000-0002-7611-9249]{Hossein Fatheddin}
\affiliation{Department of Physics, Isfahan University of Technology, Isfahan 84156-83111, Iran}
 
\author[0000-0002-0167-3595]{Sedighe Sajadian}
\affiliation{Department of Physics, Isfahan University of Technology, Isfahan 84156-83111, Iran}

\begin{abstract}
It is plausible that most of the Stars in the Milky Way (MW) Galaxy, like the Sun, consist of planetary systems, instead of a single planet. Out of the estimately discovered 3,950 planet-hosting stars, about 860 of them are known to be multiplanetary systems (as of March, 2023). Gravitational microlensing, which is the magnification in the light of a source star, due to a single or several lenses, has proven to be one of the most useful Astrophysical phenomena with many applications. Until now, many extrasolar planets (exoplanets) have been discovered through binary microlensing, where the lens system consists of a star with one planet. In this paper, we discuss and explore the detection of multi-planetary systems that host two exoplanets via microlensing. This is done through the analysis and modeling of possible triple lens configurations (one star and two planets) of a microlensing event. Furthermore, we examine different magnifications and caustic areas of the second planet, by comparing the magnification maps of triple and binary models in different settings. We also discuss the possibility of detecting the corresponding light curves of such planetary systems with the future implementation of the Nancy Grace Roman (\textit{Roman}) Space Telescope and its Galactic Time Domain survey.
\end{abstract}

\section{Introduction} \label{sec:intro}
The first discovery of a planet around a solar-type star came in 1995, when a Jovian-mass planet orbiting the star 51 Pegasi was detected from observations of periodic variations in the radial velocity of the star \citep{1995Mayor}. Since then, the exoplanet research has evolved dramatically, involving a broad range of observational and theoretical effort. 

\noindent Due to the lack of proper observational instruments and the difficulties of directly imaging exoplanets, most of these objects are detected indirectly \citep{2014Fischer} mostly with the following methods \citep[see][]{2013Wright}: Astrometric and Doppler Shift measurements (which are based on the perturbations of the radial velocity or proper motion of the host star), Timing (when a planet changes the priodicity in the brightness of a variable star), Transits (which is based on the change in the photometric light curve of a star, caused by a passing planet) and Microlensing.

\indent The \textit{Kepler}/\textit{K2} space missions for detecting exoplanets with the transit method have shown that it is quite common for a star to host multiple planets and many multi-planetary systems have been discovered this way \citep{2019Maltagliati, 2019Livingston}.

\indent Since our discussion of detecting multi-planetary systems is mainly based on the microlensing method, we will review some of its principles and foundations, in the following passages.

Since their first discovery in 1993 \citep{1993Alcock, 1993Aubourg, 1993Udalski}, the gravitational microlensing events have provided astronomers and astrophysicists with a powerful tool for studying various objects, ranging from exoplanets to Black holes. These events act as natural telescopes that can magnify and lead to the detection of some of the darkest objects in our universe, such as exoplanets orbiting the lens objects \citep{Bennett1996,Bennett2008,1991MAO,Gould1992}, Massive Astrophysical Compact Halo Objects (MACHOs) \citep{1997Alcock, 2000Alcock}, Free Floating Planets (FFPs) \citep{2021Sajadian, 2014Sumi} and even Isolated Black Holes in the Galactic disk \citep{2022Sahu, 2023Sajadian}.

The properties of microlensing events due to single and binary lenses have been studied in great details \citep[see][]{1986Schneider, 1964Liebes, 1964Rafsdal, 1966Rafsdal}. Currently, most of the exoplanets that are discovered via microlensing, belong to the category of binary events where a source star passes from behind two lenses; one star with its companion planet. The only change from the simple case of a single lens is that the deflection angle, which is caused by the gravitational field of the lens objects \citep{1920Dyson}, consists of the sum of two point lenses:
\begin{eqnarray}
\label{eq:Lens Eq b}
\alpha(\textbf{x})=\frac{4G}{c^2}(\frac{m_1(\textbf{x}-\textbf{x}_1)}{(\textbf{x}-\textbf{x}_1)^2}+\frac{m_2(\textbf{x}-\textbf{x}_2)}{(\textbf{x}-\textbf{x}_2)^2})
\end{eqnarray}
\noindent here $m_1$, $m_2$ are the masses of the two lenses and $\textbf{x}_1$,$\textbf{x}_2$ are their positions. Here, we assume two lens objects are in the same distances from the observer.  

\noindent The binary lens equation, in its complex notation, was derived by \citet{Witt1990}:
\begin{eqnarray} \label{eq:lenseq1}
\zeta=z+\frac{\mu_{\rm 1}}{\bar{ z_{\rm 1}}-\bar{z}}+\frac{\mu_{\rm 2}}{\bar{z_{\rm 2}}-\bar{z}}, 
\end{eqnarray}
\noindent where, $\mu_{1}=m_{1}/(m_{1}+m_{2})$, $\mu_{2}=m_{2}/(m_{1}+m_{2})$, $z_1$ and $z_2$ are the positions of the lenses, $\zeta=\xi+i\eta$ and $z=x+iy$ are the source and the image positions projected on the lens plane and normalized to the Einstein radius (the radius of the images ring at the complete alignment). $\bar{z}$ indicates the complex conjugate of z.  
\noindent By doing some calculations \citep{WittMao1995}, the lens equation (eq. \ref{eq:lenseq1}) becomes a lengthy fifth order polynomial in z, which requires numerical solutions in order to trace the position of the images \citep{2022Fatheddin}.

\noindent The binary microlensing method for detecting exoplanets relies on the existence of one planet in the lens system, but it is rare for a star to have only one planet. The complicated degeneracies, lack of fast and accurate models and lack of observational data, might be some of the causes for not detecting as many triple microlensing events as the binary ones (we discuss and review the probabilities of triple lens events in more details at the end of Section~\ref{sec:Triple}.).

\noindent The first observation of triple-lens gravitational microlensing was done in 2006, when a gas giant was detected orbiting a stellar binary system \citep{2020Bennett}. The first triple-lens system consisting of one star and two planets, was detected in 2008 \citep{2008gaudi}, when a Jupiter/Saturn planetary system Analog was discovered. Since then, a few other triple-lens systems, similar to these events, have been observed.

There are many lensing and microlensing projects that produce TBytes of data i.e., the Optical Gravitational Lensing Experiment (OGLE, \citet{OGLE_IV}), the Microlensing Observations in Astrophysics (MOA, \citet{MOA_gourp}) microlensing group, and the Korea Microlensing Telescope Network (KMTNet, \citet{KMTNet2016}). And the future implementation of ground and Space based surveys, like the \textit{The Nancy Grace Roman Space Telescope} (\textit{Roman}) telescope \citep{Penny2019} will produce even more data for triple-lens events. 

In this paper, after discussing the properties of triple lens systems in Section \ref{sec:Triple}, we will analyze and study the properties of magnification maps of these systems in Section \ref{sec:MagMap}. In Section \ref{sec:Roman}, we investigate the possibility of detecting the light curves of multi-planetary systems with the \textit{Roman} space telescope. Finally, in the last section, we discuss and summarize the results.
\section{Microlensing from Triple Lens Systems} \label{sec:Triple}
In this section we are going to review some characteristics of microlensing due to a triple lens system and its lens equation so that we can analyze the features and detection of multi-planetary systems in the next sections. The existence of a third lens object can cause several degeneracies in the lens equation, we review some of these degeneracies in subsection \ref{subsec: degn}. Since, most of the exoplanets that are detected with microlensing, belong to the category of binary lenses, we review some of the probabilistic comparisons between these events and events with triple lenses in the last part of this section.  
\subsection{Parameters of a Triple Lens system} \label{sec:tparam}
When a source passes from behind three lenses, the only change from a binary lens situation (equation~\ref{eq:Lens Eq b}) is that the deflection angle now consists of the sum of three point lenses:
\begin{eqnarray}
\label{eq:Lens Eq}
\alpha(\textbf{x})=\frac{4G}{c^2}(\frac{m_1(\textbf{x}-\textbf{x}_1)}{(\textbf{x}-\textbf{x}_1)^2}+\frac{m_2(\textbf{x}-\textbf{x}_2)}{(\textbf{x}-\textbf{x}_2)^2}+\frac{m_3(\textbf{x}-\textbf{x}_3)}{(\textbf{x}-\textbf{x}_3)^2})
\end{eqnarray}
\noindent here $m_1$, $m_2$, $m_3$ are the masses of the three lenses and $\textbf{x}_1$,$\textbf{x}_2$, $\textbf{x}_3$ are their positions.

\noindent In gravitational lensing formalism, the Einstein radius, $\theta_{\rm E}$, which is the angular radius of the images ring at the time of complete alignment, is defined as:
\begin{eqnarray}
\theta_{\rm E}=\sqrt{\frac{4~G~M_{\rm l}}{c^{2}} \frac{D_{\rm ls}}{D_{\rm l}~D_{\rm{s}}}}
\end{eqnarray}

\noindent where, $M_{\rm l}$ is the lens mass,  $D_{\rm{ls}}=D_{\rm s}-D_{\rm l}$, $D_{\rm l}$ and $D_{\rm s}$ are the lens and source distance from the observer.

\begin{figure}[]
\includegraphics[angle=0,width=0.49\textwidth,clip=]{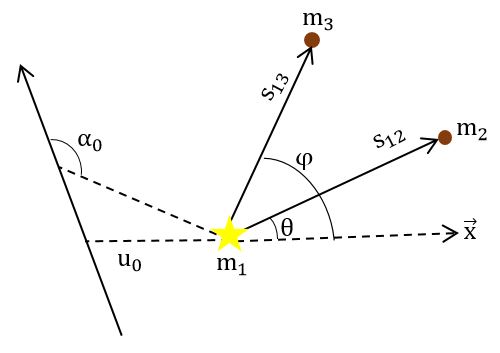}\label{fig1}
\caption{Parameters of a Triple Lens system. The three lenses are labeled as $m_1$, $m_2$, $m_3$ (where $m_1$ is the host star and $m_2$ and $m_3$ are the planets). The trajectory of the source is also illustrated. The coordinate system is centered on the host star.} \label{fig:triple}
\end{figure}

\noindent In a triple lens scenario, we use the following parameters to analyze or model an event (figure~\ref{fig:triple}):
\begin{itemize}
\item the mass ratios $q_{\rm 21}=m_2/m_1$ and $q_{\rm 31}=m_3/m_1$,
\item the triple separations $s_{\rm 21}$ and $s_{\rm 31}$ (projected on the lens plane and in units of the Einstein radius for the total masses; $M_{\rm l}=m_1+m_2+m_3$),
\item the angle $\phi$ from the $s_{\rm 12}$-axis to the $s_{\rm 13}$-axis counterclockwise and the angle $\theta$ from the x-axis of the chosen coordinate system to the $s_{\rm 12}$-axis measured counterclockwise (also in the Einstein Radius),
\item Radius of the source $\rho_{\rm *}$ (also projected in the lens plane and in the Einstein Radius),
\item the impact parameter $u_0$,
\item the angle $\alpha_0$ from the $s_{\rm 12}$-axis to the source trajectory counterclockwise,
\item time of the closest approach $t_0$ and the Einstein crossing time $t_{\rm E}$ .
\end{itemize} 
Changing any of these parameters allows for a very large variety of triple lens light curves and makes triple lenses more complicated than binary ones, which were briefly reviewed in the introduction.

The triple lens equation for a triple lens system can be derived easily as (similar to equation~\ref{eq:lenseq1}):
\begin{eqnarray} \label{eq:lenseqT}
\zeta=z+\frac{\mu_{\rm 1}}{\bar{ z_{\rm 1}}-\bar{z}}+\frac{\mu_{\rm 2}}{\bar{z_{\rm 2}}-\bar{z}} + \frac{\mu_{\rm 3}}{\bar{z_{\rm 3}}-\bar{z}}
\end{eqnarray}
Where $\mu_{i}=m_{i}/M_{\rm l}$, $z_1$, $z_2$ and $z_3$ are the positions of the lenses projected on the lens planet normalized to the Einstein radius, $\zeta=\xi+i\eta$ and $z=x+iy$ are the source and the image positions and $\bar{z}$ indicates the complex conjugate of z.

\noindent According to \citet{2002Rhie}, a polynomial form for this equation can be derived by choosing a coordinate system in a way that the position of a mass $m_{\rm 1}$, $z_{\rm 1}$, and the center of mass of the other two masses, defined as $z_{\rm 4}$, make the lens axis along the real axis of the complex plane. This process leads to a lengthy tenth-order polynomial in z, which can not be solved analytically. 

One way to solve the lens equation and calculate the corresponding light curves of a triple lens system is the Inverse Ray-Shooting (IRS) method \citep{1986Kayser,1990Wambsganss,1999Wambsganss}, discussed in Appendix \ref{sec: IRS}. One method to increase the performance speed of IRS method is using tree algorithm as explained in \citet{2010sajadian}.

\indent In the next subsection, we will review some degeneracies in the triple lens equation.    
\subsection{Degeneracies in Triple Lens Equation} \label{subsec: degn}
There are many parameters involved in the shapes of light curves for  microlensing events and consequently, they can have different shapes. Sometimes, this can lead to a parameter degeneracy in the modeling of the event either accidentally \citep[see, e.g., ][]{1998ApJGaudi} or due to symmetry of the lens equation \citep[see, e.g., ][]{1994ApJGould}.

\indent In the case of triple microlensing, a four-fold degeneracy in the light curves can be found, which is similar to the case of close and wide degeneracy found in a binary lens \citep{2014Song}. Three important degeneracies are expected for triple lens microlensing events (which are discussed in detail in \cite{2014Song}.): Discrete degeneracy, Four-fold close/wide degeneracy and Continuous external shear degeneracy. Below, we briefly review each degeneracy.

\noindent Discrete degeneracy arises if one reverses the sign of the three parameters $(u_{\rm 0}, \alpha_{\rm 0}, \phi)$, which  is owing to the symmetry of the lens equation. This degeneracy can be resolved by measuring the second-order parallax effect. In planetary microlensing, where the mass ratios are $q_{21}\ll1$ and $q_{31}\ll1$, one can denote the planetary positions in complex notation (equation \ref{eq:lenseqT}) as $z_{\rm p}$. If one changes $z_{\rm p}$ into $z^{-1}_{\rm p}$, the shape of the central caustics remains the same. This degeneracy can also be found in the case of binary lenses \citep{1999DominikAA} and it is not just limited to triple lenses.

\noindent Finally, if one rewrites the triple lens equation and expands the deflection terms caused by two of the masses ($m_2$ and $m_3$) at the location of the primary mass ($m_1$) using the Taylor expansion, we end up arriving at a lens equation similar to the case of binary lenses (equation~\ref{eq:lenseq1}). The important consequence of this continuous degeneracy is that in some cases, multi-planetary systems may be mistakenly detected as a single-planet system \citep{2014Song}. 
  
\subsection{Detection Probability of Triple Lensing Events} \label{sec:prob}
For detecting an exoplanet via microlensing, a signal is considered planetary, if the deviations between the light curve calculated by the binary lens model (where one of the lens objects is planetary) and the simple single lens model is larger than a carefully chosen critical value. The Gould \& Loeb criterion \citep{Gould1992} provides a reliable criterion by considering deviations as planetary signals when a few observational points deviate consecutively from the single-lens light curve.

 \noindent \cite{2011Ryu} show that if we consider the Gould \& Loeb criterion for detecting a planet, for most cases, the detection probability of the low-mass planet in the triple lens system is very similar to that in the binary lens system. The reason for this similarity is due to the lens superposition in the central regions \citep{2001Han}. It should also be noted that if the interference between caustics of the two planets change the original shapes, the binary superposition would not work well.

 \indent Additionally, it can be demonstrated that for the low-magnification events, where the source star only passes one of the planetary caustics, the probability of detecting the second planet is very low \citep{2002Han}. But, if the source star passes the main caustic (i.e. the high-magnification events), the probability of detecting the second planetary companion increases dramatically to unity \citep{1998Griest}. This is due to the closeness of planets to the Einstein radius and makes microlensing a unique method for detecting multi-planetary systems in such situations \citep{2012Gaudi}.

\section{Microlensing Magnification Map Analysis} \label{sec:MagMap} 

Magnification maps are 2-dimensional planes that show the magnification in the light of the source star in a region of the source plane. In a magnification map, each point is a solution of the lens equation (equation \ref{eq:lenseqT}) regardless of the source trajectory. Therefore, magnification maps are helpful tools for discussing and analyzing microlensing events and their magnification.

\indent In this section we analyze different configurations of the triple lens magnification maps by evaluating the change in average magnification due to the second planet when the corresponding parameters of the system change and evolve.

\indent The magnification due to lens objects from a point-like source star in any coordinate, $\textbf{x}$, of the source plane can be easily calculated by the inverse of the determinant of amplification matrix \citep{1992Schneider}:
\begin{eqnarray}
\label{eq:mag}
\mu(\textbf{x})=\frac{1}{det A(\textbf{x})}
\end{eqnarray}

\noindent where the amplification matrix is:
\begin{eqnarray}
\label{eq:amp}
A(\textbf{x})=\textbf{1}-\frac{\partial {\bf \alpha} (\textbf{x})}{\partial \textbf{x}}
\end{eqnarray}

\indent If we input deflection angles due to binary (equation \ref{eq:Lens Eq b}) and triple lenses (equation \ref{eq:Lens Eq}) in equation \ref{eq:mag} and consider the source plane as a $L_x\times L_y$ rectangular plane consisting of $N_x\times N_y$ points, i.e. pixels, we get triple lens and binary lens magnification maps due to a point-like source star, respectively. 

Furthermore, the average magnifications due to binary and triple lensing systems can be derived similarly as: 
\begin{eqnarray}
\label{eq:mut}
\left<\mu\right>=\frac{1}{N_x\times N_y}\sum_i^{N_x} \sum_j^{N_y}\mu(x_{ij})
\end{eqnarray}

\noindent Now, the magnification due to the second planet can be defined as the deviation between the mean magnifications of triple, $\left<\mu_t\right>$, and binary systems, $\left<\mu_b\right>$,:
\begin{eqnarray}
\label{eq:mut}
\mu_3=\left<\mu_t\right>-\left<\mu_b\right>
\end{eqnarray}

\indent There are some regions of the magnification map that the magnification is very high (almost infinite), i.e. the caustics \citep{1993Erdlschneider}. Caustics can form due to the host star and the planets as well. Studying the properties of the planetary caustics is important because the characteristics of the corresponding planetary signals in the microlensing light curve depend highly on them \citep{2006Han}.

\begin{figure}[]
\includegraphics[angle=0,width=0.49\textwidth,clip=]{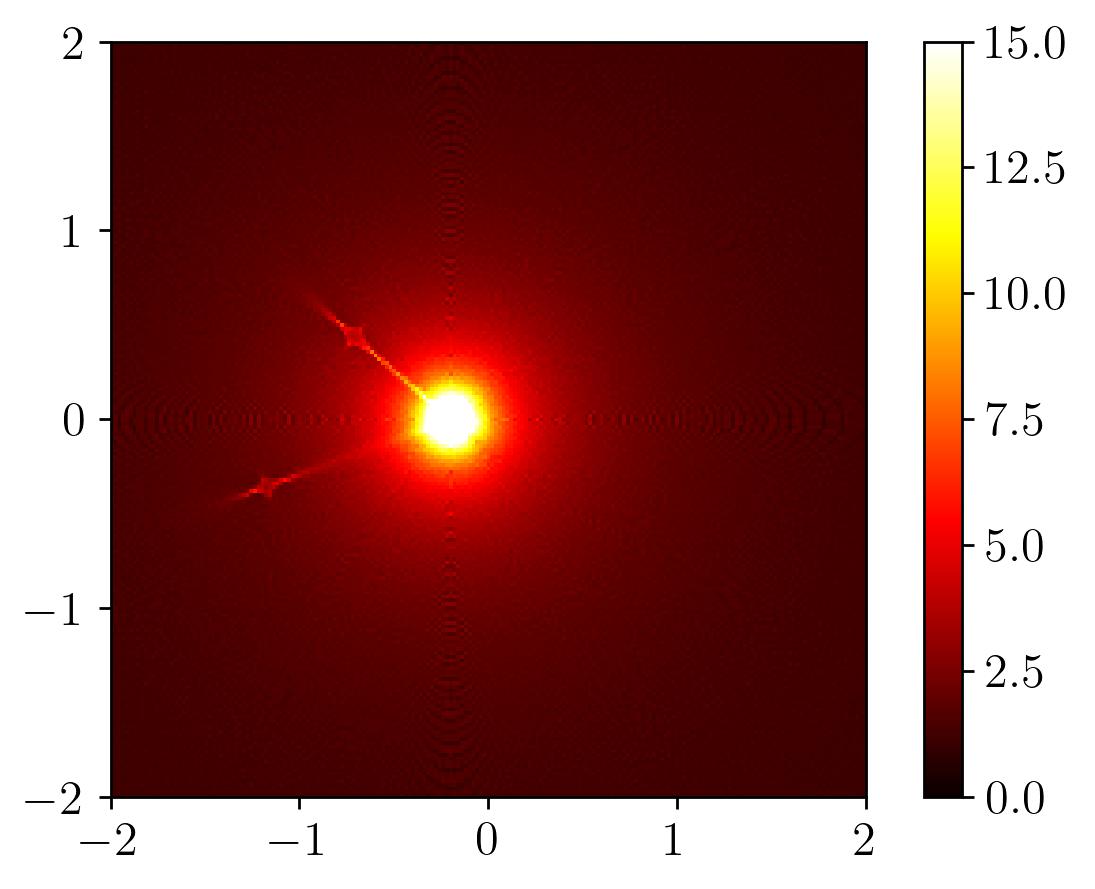}	
\caption{The orbital motion of two planets around the host star and the change in the system's magnification map. The color bar displays the mean magnification at each point in the source plane. An animated version of this figure is available in the HTML version of the final article. The change in the respective planetary caustics can be seen.}
\label{fig:orbit}
\end{figure} 

\noindent One of the most important properties of planetary caustics is their size or area; since it is used to determine the probability of detecting the planet. If the planetary caustic is big enough, there is a higher chance of a source star passing through, or close, to it. Which, in turn, leads to a more detectable planetary signal.

\indent It is known that a planet and its host star engage in a Keplerian motion, where they move in elliptic orbits around their common center of mass. This motion effects the magnifications due to each planet and the size of their caustics. However, if the xallarap effects can be neglected \citep{2021Rota}, a planetary signal in the microlensing light curve only represents a snapshot of the planet at its current angular separation from the host star, and does not depend on its orbital motion \citep{2009Rahvar}.

 \begin{figure}[]	
	\subfigure[]{\includegraphics[angle=0,width=0.49\textwidth,clip=]{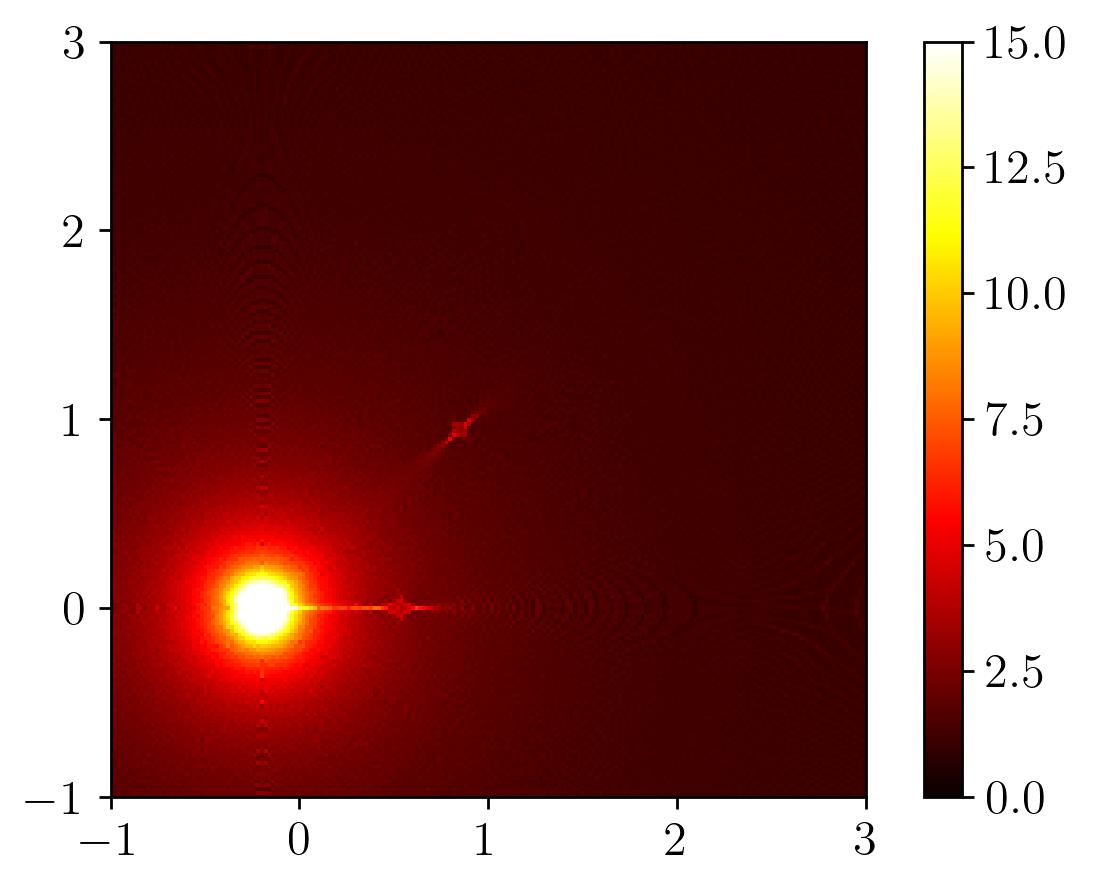}\label{fig3}}		
	\subfigure[]{\includegraphics[angle=0,width=0.49\textwidth,clip=]{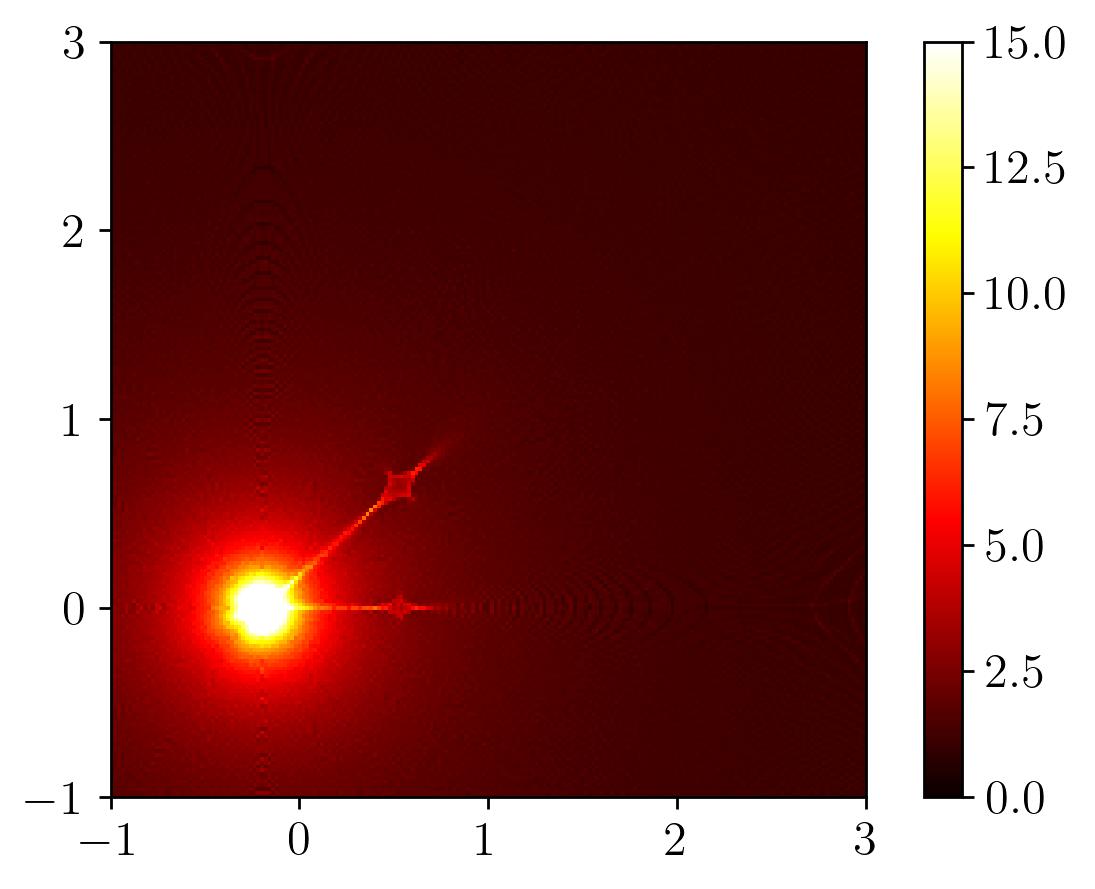}\label{fig4}}			
	\caption{Top panel: This animation illustrates the dynamic change in the magnification map of the system as the separtion of the second planet from the host star ($s_{31}$) changes. Bottom panel: This animation shows that dynamic change in the magnification map of the system as the mass ratio of the second planet ($q_{31}$) changes. An animated version of this figure is available in the HTML version of the final article. \label{fig:mma}}
\end{figure} 

\noindent Figure \ref{fig:orbit} shows an animation of the change in the magnification map of a multi-planetary system which is caused by the orbital motions of two planets around the host star. It can be seen that the sizes of the planetary caustics change as the planets rotate around the barycenter. Additionally, the color of the caustics also changes, which represents the variation in the mean magnification of each caustic. 

\indent As it was mentioned in the introduction, the case of gravitational microlensing due to binary lenses has been studied extensively before. Also, In subsection \ref{sec:tparam}, the main parameters of a triple lens system were reviewed. Some of these parameters are only introduced in the case that the third lens (the secondary planet) exists. These parameters are: $(q_{31}, s_{31},\phi)$. 

\noindent In order to measure and discuss the changes in $\mu_3$, we assume that the parameters of the primary planet are constant and only those of the second planet vary. The change in the $\phi$ angle was demonstrated in Figure \ref{fig:orbit}.

\noindent In order to investigate the dependence and change of $\mu_3$ with its angular separation $s_{31}$ and mass ratio $q_{31}$, we consider the position of the first planet constant at $\theta=0.0$ degree, $s_{21}=1.0~\theta_{E}$ with the mass ratio $q_{21}=9\times10^{-4}$ (which is about the mass ratio of the planet Jupiter and Sun). The angle $\phi$ of the second planet is also considered constant at $60$ degrees. 
  
 \begin{figure*}[]	
	\subfigure[]{\includegraphics[angle=0,width=0.49\textwidth,clip=]{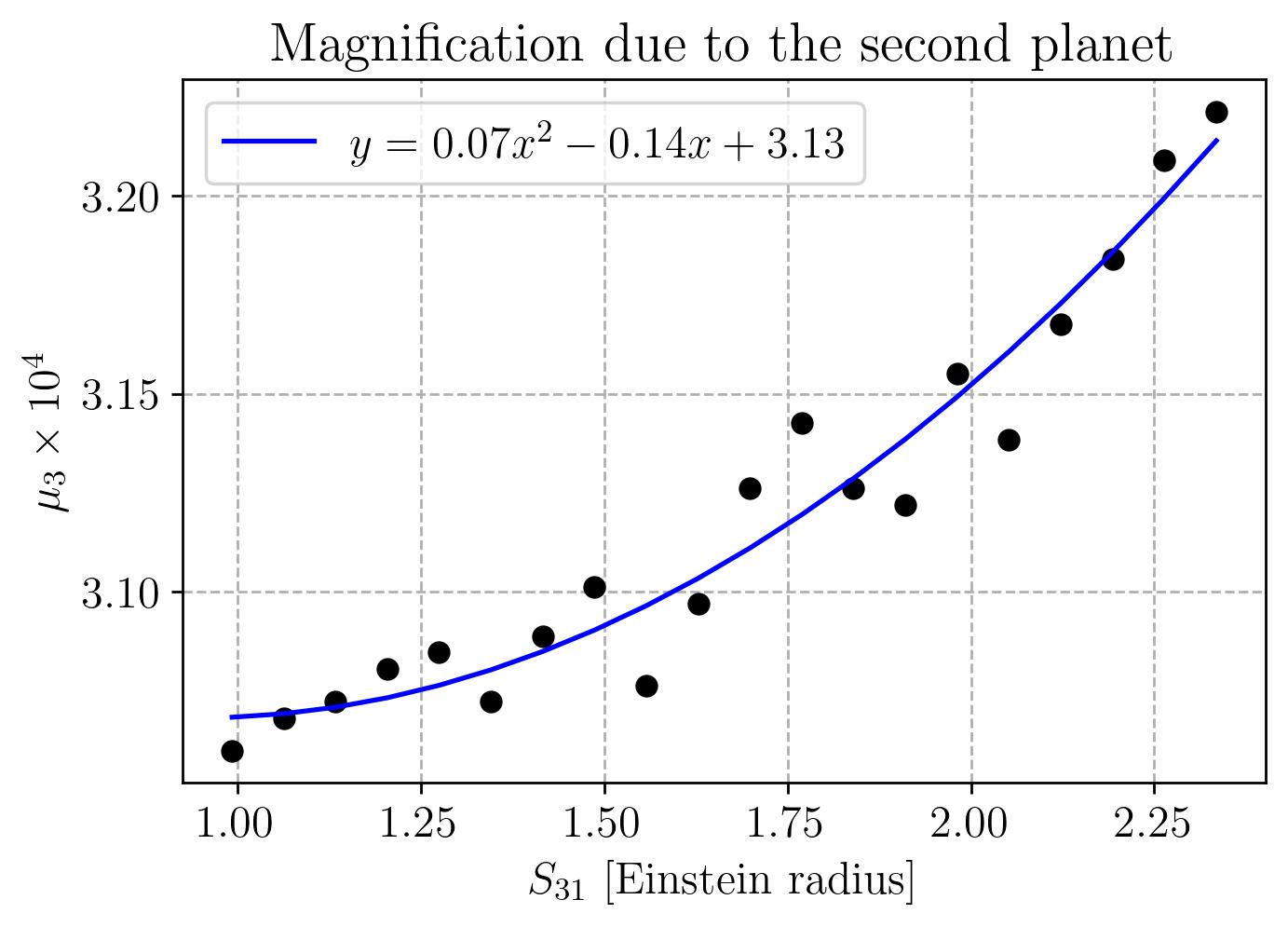}\label{fig5}}		
	\subfigure[]{\includegraphics[angle=0,width=0.49\textwidth,clip=]{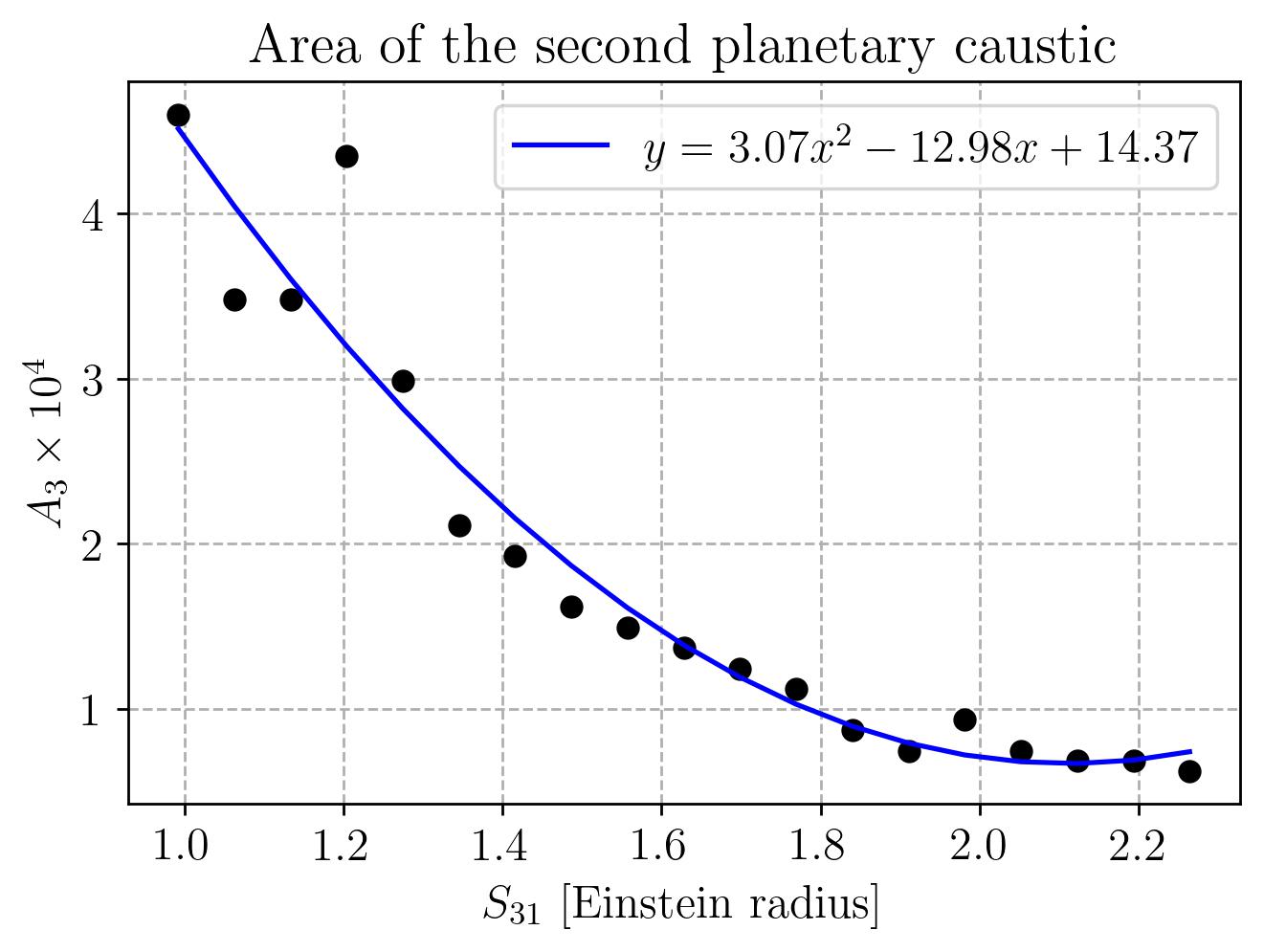}\label{fig6}}		
	\subfigure[]{\includegraphics[angle=0,width=0.49\textwidth,clip=]{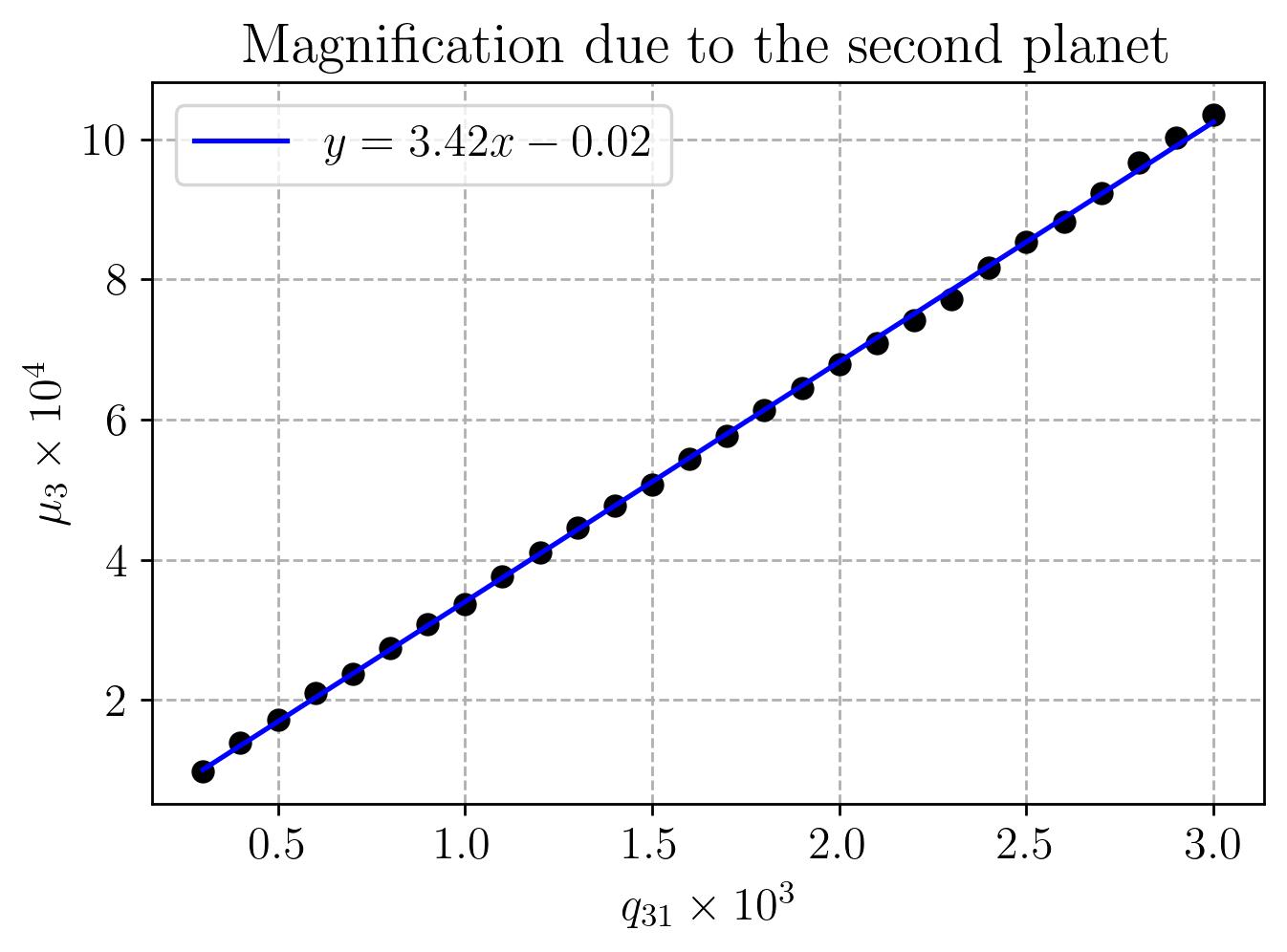}\label{fig7}}	
	\subfigure[]{\includegraphics[angle=0,width=0.49\textwidth,clip=]{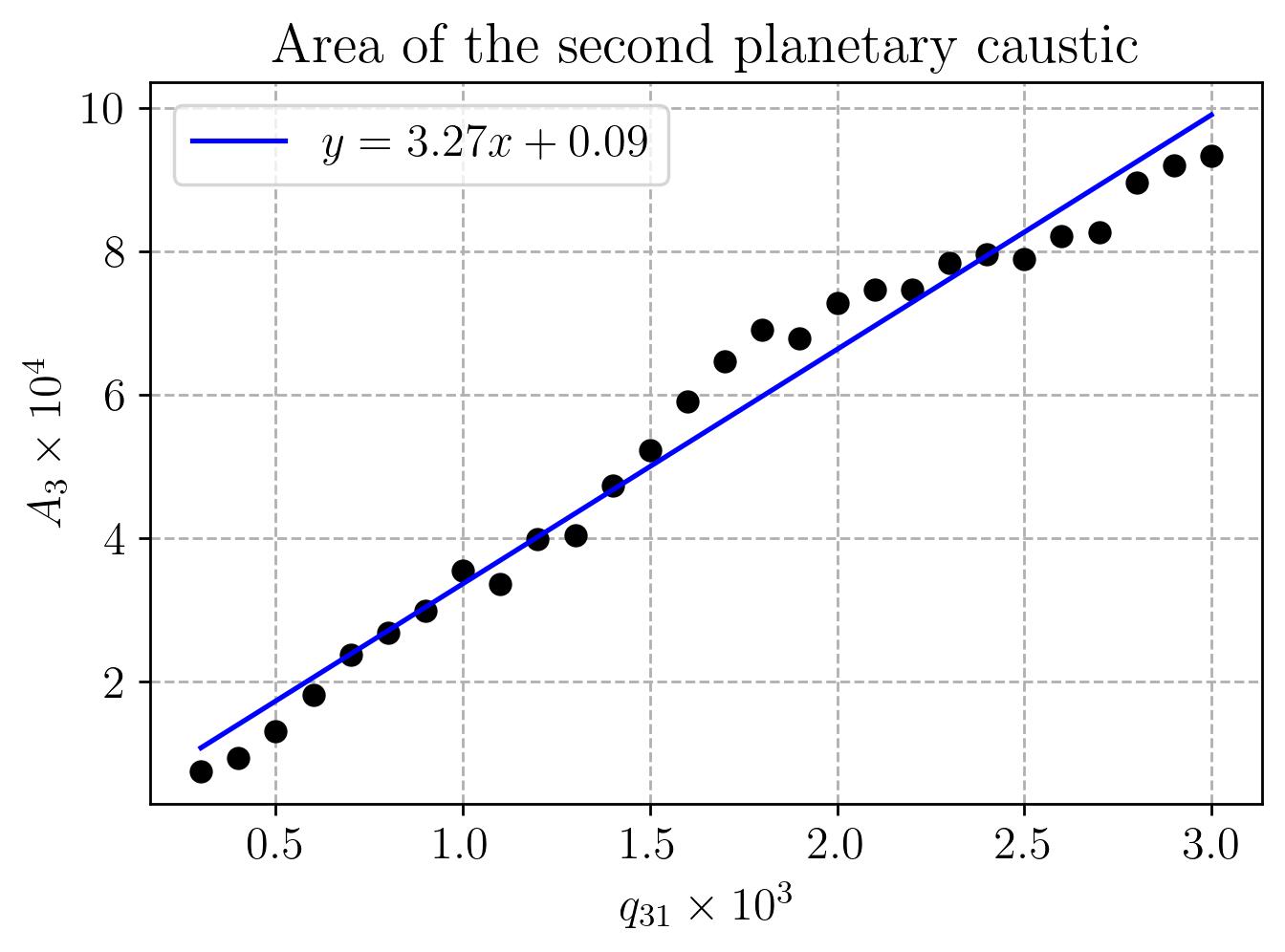}\label{fig8}}		
	\caption{Figure (a) and (b) show The change in the mean magnification and caustic area of the second planet with respect to its separation ($s_{31}$). Figure (c) and (d) show the change in the mean magnification and caustic area of the second planet with respect to its mass ratio ($q_{31}$). \label{fig:results}}
\end{figure*} 

\indent First, we consider $s_{31}$ varying between $0.9~\theta_{E}$ and $2.5~\theta_{E}$ where the second planet is also a Jovian planet with the mass ratio $q_{31}=9\times10^{-4}$. Figure \ref{fig3} shows the dynamic change in magnification and area of the second planetary caustic with respect to its separation from the host star. We can see that the second planetary caustic becomes brighter (which represents the increase in its average magnification) and smaller as the second planet moves away from the star. The change in $\mu_3$ with separation $s_{31}$ and the best-fitted curve are shown in Figure \ref{fig5}. It can be seen that $\mu_3$ increases with $s_{31}$. However, some deviations from the best fit curve are visible in this figure, which are mostly caused by the approach and retreat of the second planet to the primary planet. It should also be noted that, although the mean magnification increases with distance, but the area of the caustic decreases considerably, which is shown in Figure \ref{fig6}.

\indent Then, we assume that the position of the second planet remains constant at $s_{31}=1.1~\theta_{E}$ and the mass ratio, $q_{31}$ changes between $3\times10^{-4}$ and $3\times10^{-3}$. The dynamic changes of the second planetary caustic are illustrated in Figure \ref{fig4}. We can see that the second planetary caustic becomes brighter and larger (therefore, more detectable) as $q_{31}$ increases. Figure \ref{fig7} demonstrates the change in $\mu_3$ as $q_{31}$ increases. We can see that $\mu_3$ increases linearly with the best fit line: $y=3.42x-0.02$. The change in the area of this caustic is also shown in Figure \ref{fig8}, which shows that the caustic area (and the detection probability) of the second planet also increases linearly with respect to $q_{31}$.

\indent The animations used in this section can also be found at: \href{https://github.com/AstroFatheddin/Triple-Lens-Animations}{https://github.com/AstroFatheddin/Triple-Lens-Animations}

\section{The \textit{Roman} Galactic Time Domain Survey} \label{sec:Roman}
The \textit{Roman} space telescope's Galactic Time Domain Survey (RGTDS) will detect an unprecedented number of exoplanets by monitoring the light curves of a large sample of stars toward the center of our galaxy. RGTDS is expected to discover bound exoplanets where the planets have distances in the range of 1 au and more from their host star using the gravitational microlensing \citep{Penny2019}. Another possible discoveries by the \textit{Roman}\ telescope are detection of habitable exoplanets, exomoons rotating free floating planets, and exoplanets around source stars \citep{sajadian2021b,2023MNRAsang,2019MNRASbagheri}   

\indent In order to explore the detection of multi-planetary systems in RGTDS, we need to simulate a sample of different triple lens planetary microlensing light curves and see if there is a significant difference between the triple and binary models that is observable by the \textit{Roman}\ telescope.   

\noindent We prepared a sample of 55 microlensing events where we considered both binary and triple systems and their difference. In the triple models, the planet that was considered in the corresponding binary systems also exists, with the addition of another planet. This way we can investigate that if a planet is detected with microlensing and the \textit{Roman}\ telescope, are the effects of the second planet also detectable in the light curve.

 \begin{figure*}[]	
	\subfigure[]{\includegraphics[angle=0,width=0.49\textwidth,clip=]{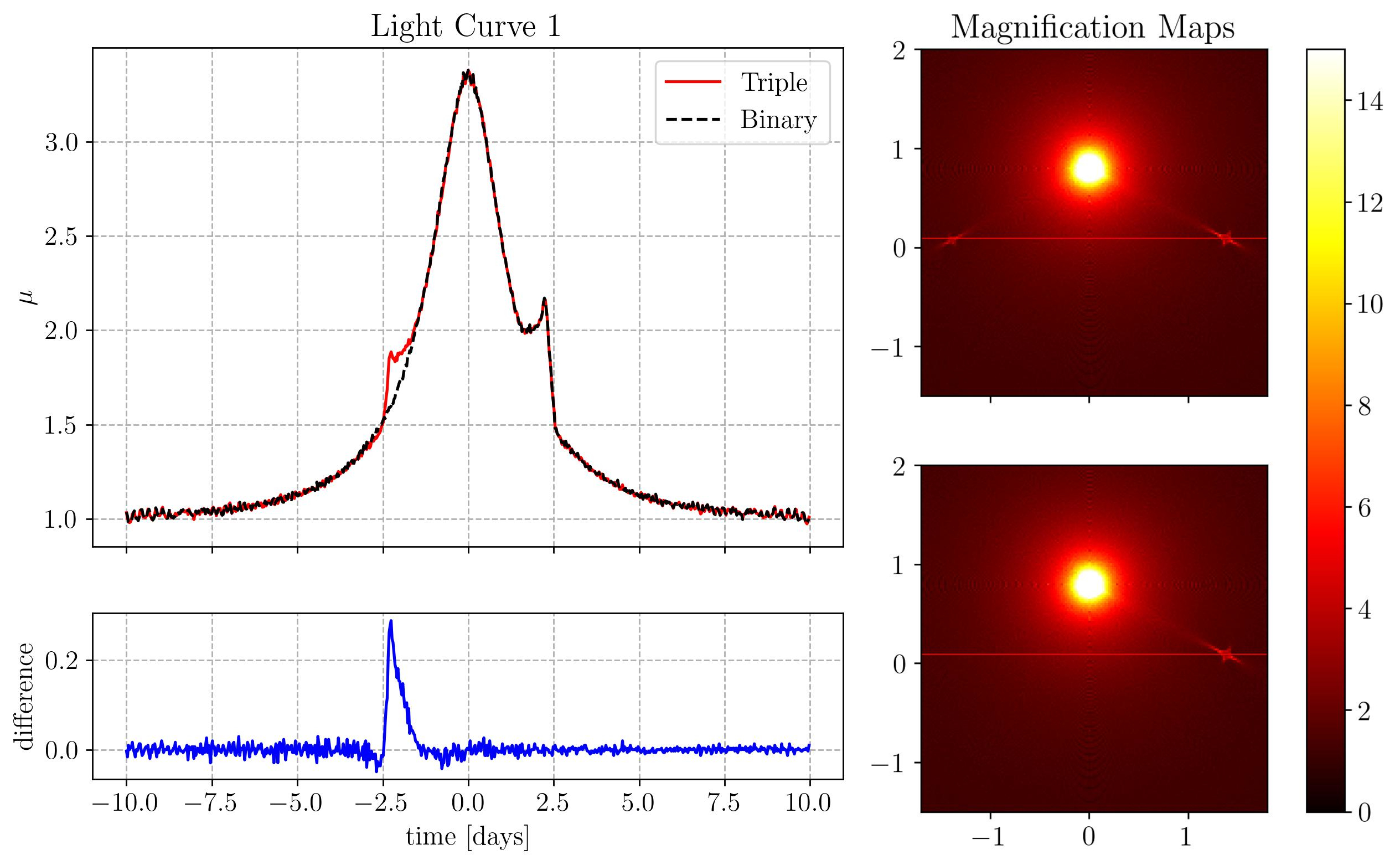}\label{fig9}}		
	\subfigure[]{\includegraphics[angle=0,width=0.49\textwidth,clip=]{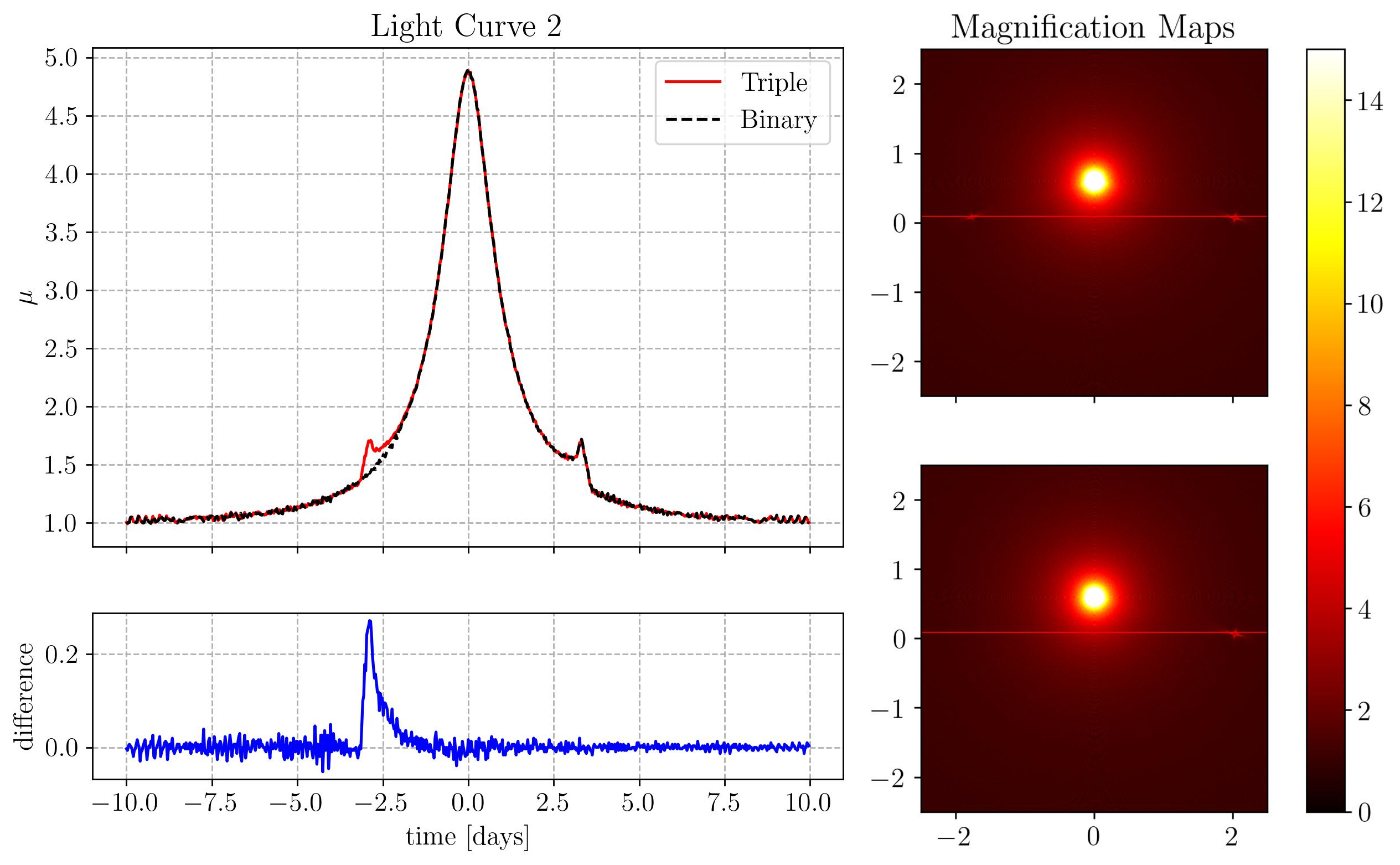}\label{fig10}}		
	\subfigure[]{\includegraphics[angle=0,width=0.49\textwidth,clip=]{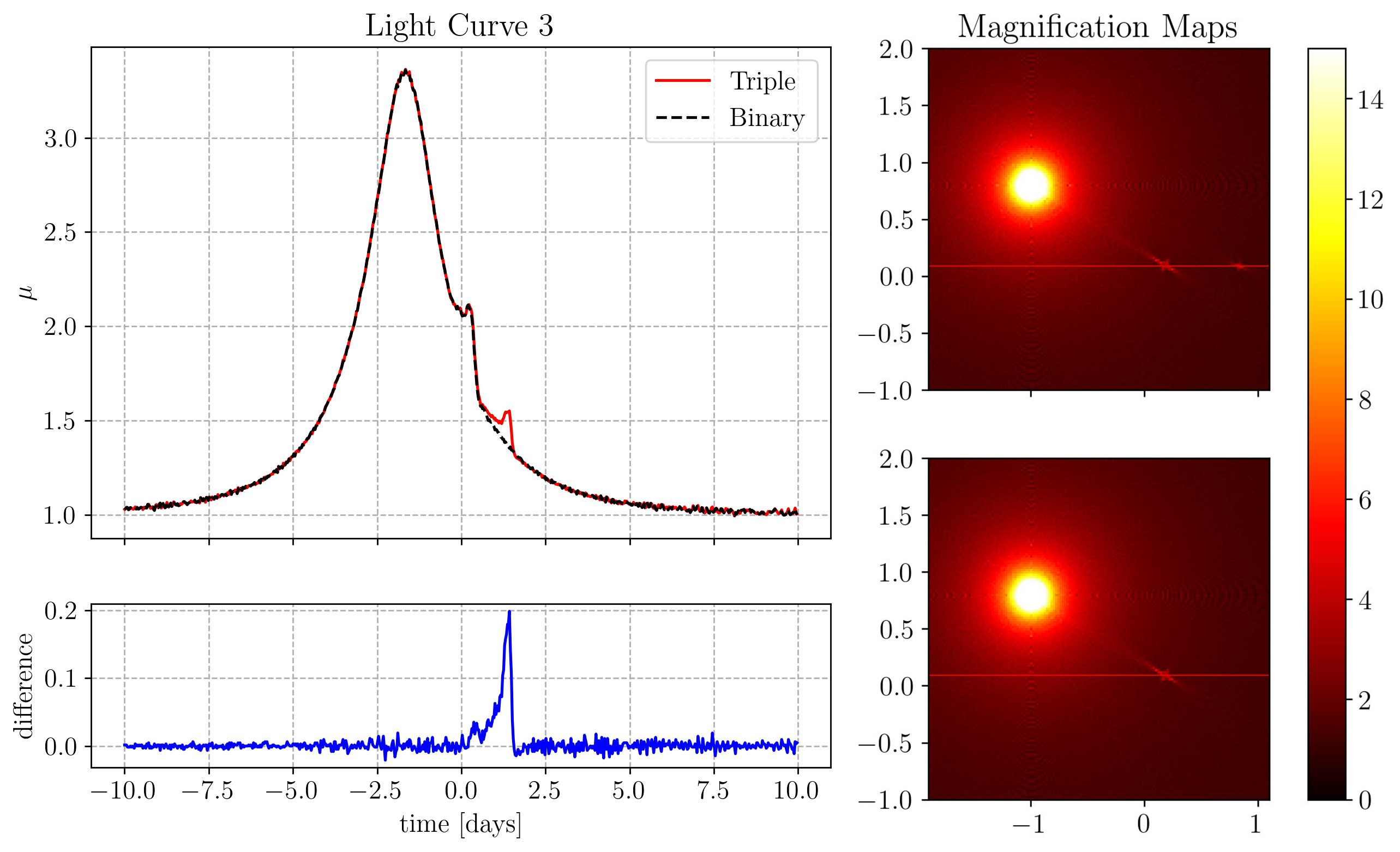}\label{fig11}}	
	\subfigure[]{\includegraphics[angle=0,width=0.49\textwidth,clip=]{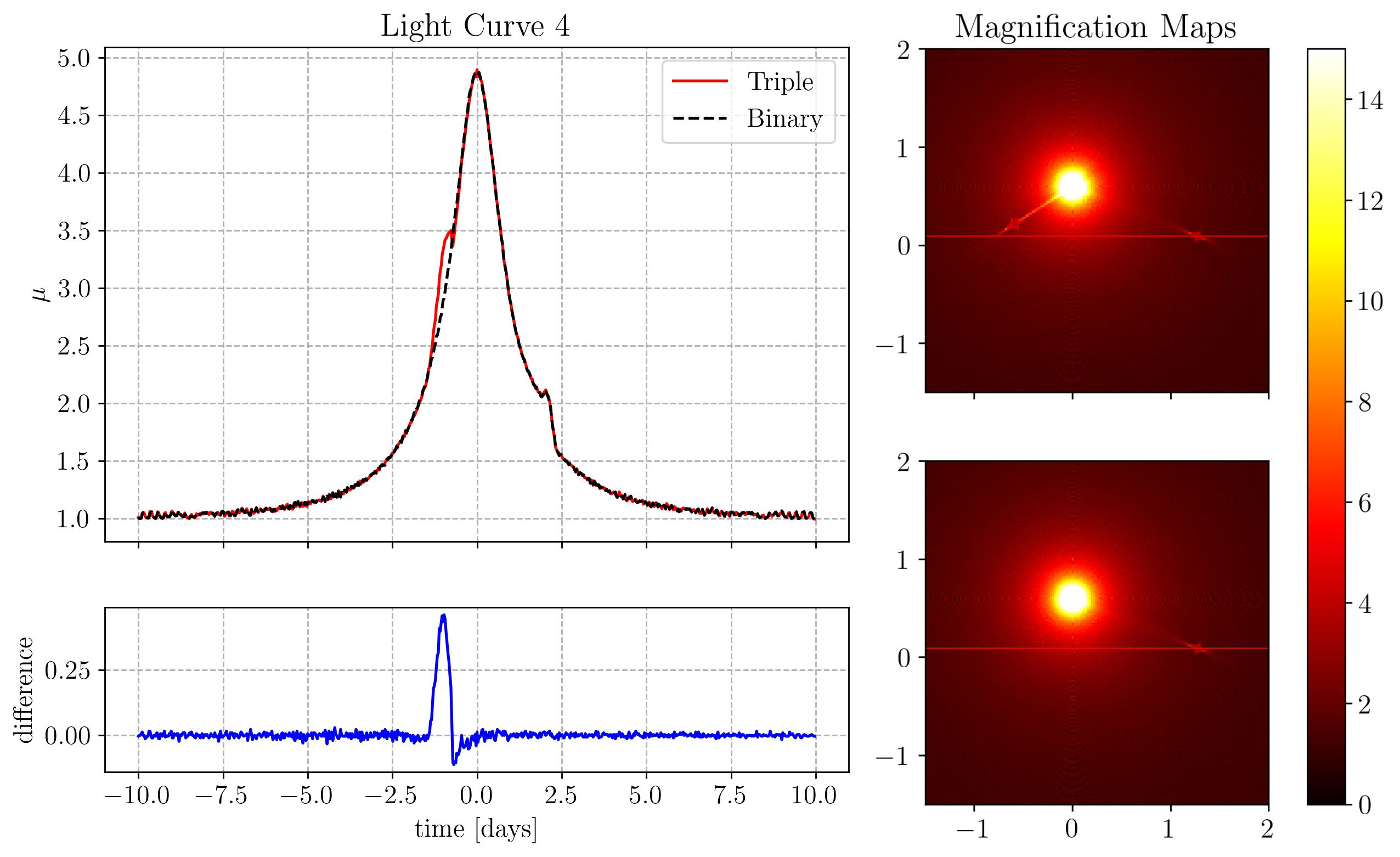}\label{fig12}}		
	\caption{Light curves of four different microlensing events. In each figure the triple and binary models and their difference are illustrated. The magnification map on the top right of each light curve corresponds to the configurations of the lenses and the path of the souce star where there is a second planet (the red curve), whereas the maginfication map at the bottom represents the binary system configuration (the dashed black curve).   \label{fig:lc}}
\end{figure*}

\indent In our sample we have considered a star with the radius $\rho_*=0.05~R_{\rm E}$ as the source star. This star moves behind the lenses with an impact parameter of $u_0=0.09$ and the trajectory angle of $\alpha_0=0.0$ degree. We take these properties of the source star constant throughout our sample and instead change the configurations and coordinates of the lens system in order to simulate a variety of different events. Also, instead of considering the position of the host star as the center of our reference frame, we make a rectangular plane with the size of $L_x\times L_y$ around the path of the source star and consider the center of this plane as our reference frame. This approach makes the light curve simulations (which are explained in appendix \ref{sec: IRS}) easier and allows for a wide range of samples by just moving the positions of the lens objects with respect to the origin.

\indent Four light curves, taken from the sample, are shown in Figure \ref{fig:lc}. The light curves corresponding to triple systems are shown with the red curve, whereas the binary system light curves are illustrated with a dashed black curve. The difference between the triple and binary systems are displayed with a blue curve at the bottom of each light curve. We can see that the planetary signal due to the second planet is clearly visible in every light curve. In each light curve, $q_{12}$ is $0.002$, $0.002$, $0.0009$, $0.001$ and $q_{13}$ is $0.001$, $0.001$, $0.0005$, $0.0009$ respectively.

\indent To simulate the synthetic data points taken by the \textit{Roman} telescope, we assume these observations are done in the W149 Filter which is a combination of the stellar magnitude in the standard bands. The cadence between data is fixed at $15.16$ min. \textit{Roman}\ will detect the Galactic bulge during six $62$-day observing seasons. Three seasons will be done during two first years of the \textit{Roman}\ mission, and others will happen during two last years of its mission. So there is a long gap between first three observational seasons and others.  

\noindent In fact, the \textit{Roman}\ telescope can detect the Galactic bulge only during two 62-days time intervals in a year. Hence, the time interval between two observing seasons will be at least $120$ days. More details about these simulations can be found in \citet{sajadian2021b}.

\indent As it was mentioned in subsection \ref{sec:prob}, in planetary microlensing with one planet, the difference between the $\chi^2$ values derived from fitting the real binary model, $\chi^2_{binary}$, and the best-fitted single lens model, $\chi^2_{single}$ is considered for justifying a detection. Here, we consider the same notion for investigating that if a second planet is detected by \textit{Roman}.\ We assume that the system is in fact triple and take the following value as our criterion for detecting the second planet:
\begin{eqnarray}
\label{eq:mut}
\Delta \chi^2=\chi^2_{triple}-\chi^2_{binary}.
\end{eqnarray}         

\noindent We have calculated the $\Delta \chi^2$ values for our sample of light curves by considering various apparent magnitudes at the baseline W149 ($m_{\rm{W149}}$), different Einstein crossing times ($t_{\rm E}$ [days]), Planetary mass ratios ($q_{12}$ and $q_{13}$) and angular separations ($s_{12}$ and $s{13}$). A statistical description of our results can be found in table \ref{tab:chi}. $Q_1$, $Q_2$ and $Q_3$ are the $25th$, $50th$ (the median) and $75th$ percentiles, respectively. 

\begin{deluxetable}{ccccc}
\tablenum{1}
\tablecaption{statistical description of the results  \label{tab:chi}}
\tablewidth{0pt}
\tablehead{\colhead{}&\colhead{mean}&\colhead{$Q_1$}&\colhead{$Q_2$}&\colhead{$Q_3$}}
\startdata
\textbf{$t_{\rm E}$} & 14.78 & 9.95 & 14.77 & 19.73  \\
\textbf{$m_{\rm{W149}}$} & 19.48 & 18.31 & 19.53 & 20.52 \\
\textbf{$q_{12}$} & 0.0018 & 0.0009 & 0.0010 & 0.0020  \\
\textbf{$q_{13}$} & 0.0011 & 0.0007 & 0.0009 & 0.0015  \\
\textbf{$s_{12}$} & 1.3645 & 1.2080 & 1.3750 & 1.5905  \\
\textbf{$s_{13}$} & 1.3944 & 1.2095 & 1.3370 & 1.5615  \\
\textbf{$\Delta \chi^2$} & 70267.02 & 242.94 & 743.49 & 9841.15 
\enddata
\end{deluxetable}

\noindent The threshold of $\Delta \chi > 500$ is usually considered as the criterion for detecting a planetary signal in the light curve of a microlensing event. If we consider this criterion for our light curve sample, 35 out of 55 events have a robust planetary signal which was detected by \textit{Roman}, which gives an efficiency of about $60\%$. It should also be noted that since many parameters are involved in the shape of a microlensing light curve (subsection \ref{sec:tparam}), our reported efficiency is also highly dependent on our chosen values for these parameters. Nevertheless, our data shows that there is a high probability of detecting multi-planetary systems with the future implementation of \textit{Roman},\ provided that the planets have the proper features for detection.

\indent The light curves data and the full simulation results of the \textit{Roman} space telescope can be found at: \href{https://github.com/AstroFatheddin/Triple-Lens-Data}{https://github.com/AstroFatheddin/Triple-Lens-Data}. 

\section{Conclusions}
\indent It is quite plausible that a star can host multiple planets, as it has been evident in the transit data from space telescope like \textit{Kepler}\ and the Transiting Exoplanet Survey Satellite (TESS). 

\indent In this work, we studied the detection and characterization of multi-planetary systems with gravitational microlensing. We have shown that different parameters can affect the shape of microlensing magnification maps and proportionately change the shape of corresponding light curves. Also, we have shown that the area of the second planetary caustic, and therefore its detection probability, are also highly dependent on its configurations (i.e. its separation from the host star or its mass ratio).
    
\indent With simulating a large variety of light curves for the triple events, we studied the possibility of detecting multi-planetary systems with the future implementation of the \textit{Roman} space telescope. Our results show that we can expect \textit{Roman} to provide a large and diverse sample of multi-planetary systems using the microlensing method.

\indent In this paper, we mainly considered the multi-planetary system consisting of a host star with two planets. However, this results can also be simply investigated for a system with a higher number of planets.

\section*{Acknowledgements}
H. F. thanks Mozafar Allahyari for his help and support. We also thank Reza Niyazi and Amirhossein Jafari for their assistance in computing some of the light curves.

\appendix
\section{Magnification Map and Light Curve Simulations} \label{sec: IRS}
Most of the simulations of binary microlensing events are modeled by using the contour integration method which is based on the Green's theorem \citep{1997Gould,1998Dominik,Bozza2010} in order to infer the characteristics and the physical parameters of the events. In this method, the total area of the lensed images is calculated, which leads to the magnification factors and light curves. The Inverse Ray-Shooting (IRS) \citep{1986Kayser,1990Wambsganss,1999Wambsganss} is not used in these cases due to its inefficiency for fitting light curves of low-multiplicity point mass lenses of microlensing planetary systems \citep[see][]{2002Rhie}. But, IRS proposes an efficient method for modeling planetary microlensing events with more than two lenses \citep{Bennett2009}. Here, we briefly explain the IRS method which was used for simulating the magnification maps of section \ref{sec:MagMap} and Light curves of section \ref{sec:Roman}.  

\indent When we observe a Gravitational Lensing or Microlensing event, the light rays are emitted from the source (which is in the source plane) and after moving through the Lens plane, they are deflected due to the gravitational field of the Lens. The special theory of relativity states that the light rays are "time-invariant", which means that the time reverse of the path that light rays take from any source to an observer is allowed. This is the main notion behind IRS.

\noindent In the formalism of IRS, for inverting the lens equation, equation \ref{eq:lenseqT}, (i.e. knowing the position, distortion, magnification, of the image(s) of the source), we shoot rays backwards from the observer to the lens plane and by calculating the deflection angle of each ray (equation \ref{eq:Lens Eq}), we find the location of where the light ray hits the source plane, and therefore forming an image there. Because of the very long distances in Astronomy, the observer can not differentiate between most close and far objects, so in IRS, everything is happening in one plane which is the observed plane.

\noindent Since the source is moving behind the Lens objects (we usually take the lens to be stationary), we can make a set of frames and in each frame propagate the rays by simulating the motion of the Source and then take a snap-shot. Finally a dynamic light curve can be made by calculating the areas of the formed images and then dividing them by the area of the unlensed source.

\indent We also note that in our simulations, in order to consider the finite source effects, we have considered the source star as an array-like homogeneous disk with the radius $\rho_*=0.05R_E$.

\indent A \textsc{Python} code for simulating the magnification maps and light curves used in this paper and based on the above discussions was developed by the authors and can be found at:

 \href{https://github.com/AstroFatheddin/LenSim }{https://github.com/AstroFatheddin/Triple-LenSim}.

\bibliography{sample631}{}
\bibliographystyle{aasjournal}
\end{document}